\documentclass[journal]{new-aiaa}

\usepackage{graphicx}
\usepackage{amsmath}
\usepackage{longtable}

\usepackage{color}
\definecolor{link}{rgb}{0.078,0.353,0.588}
\hypersetup{colorlinks=true,urlcolor=link,citecolor=black,linkcolor=black}

\newcommand{\vobj}{\mathbf{v}_{\mathrm{obj}}}
\newcommand{\vown}{\mathbf{v}_{\mathrm{own}}}
\newcommand{\vpx}{v_{\mathrm{px}}}
\newcommand{\FOV}{\mathrm{FOV}}

\title{Limits on Kinematic Inference from the PURSUE Airborne Sensor Videos
}

\author{Jacob Haqq-Misra\footnote{jacob@bmsis.org (Corresponding Author).
ORCID: 0000-0003-4346-2611.}}
\affil{Blue Marble Space, Seattle, WA 98104}

\author{Ravi Kopparapu\footnote{ravikumar.kopparapu@nasa.gov. ORCID: 0000-0002-5893-2471.}}
\affil{NASA Goddard Space Flight Center, Greenbelt, MD 20771}

\begin{document}

\maketitle

\begin{abstract}
This paper analyzes the 112 sensor videos released under the PURSUE (Presidential Unsealing and Reporting System for UAP Encounters) initiative. Redactions to on-screen displays limit information available for identifying the reported unidentified anomalous phenomena (UAP). Converting the rate at which any object transits across part or all of a frame from pixels per frame to a physical velocity requires (1) the range to the object, (2) the velocity of the observing aircraft, (3) the aspect angle between the object's path and observer, and (4) the field angle of the camera. No PURSUE video provides this set of information. In one clip (DOW-UAP-PR113), the object crosses depression-angle markings, which allows reconstruction of the camera's field of view. Even this clip does not provide sufficient information to distinguish between a bird flying at close range and a large craft traveling at Mach 5 a kilometer away. A second clip (DOW-UAP-PR149) includes a vessel of known size in the frame, which can be used as a reference to calculate an upper bound of Mach 0.4 on the object's relative velocity. Without further information, the PURSUE sensor videos in their present form cannot fully resolve unidentified cases or conclusively indicate anomalous velocities.
\end{abstract}

\section*{Nomenclature}

{\renewcommand\arraystretch{1.0}
\noindent\begin{longtable*}{@{}l @{\quad=\quad} l@{}}
$f$ & frame rate of the video container, frames\,s$^{-1}$ \\
$f_{\rm px}$ & focal length expressed in pixels, px \\
$H$ & frame height, px \\
$k$ & angular scale of the sensor at the object's position in the frame, px\,rad$^{-1}$ \\
$L$ & physical length of an in-frame reference object, m \\
$l_{\rm px}$ & image extent of the reference object, px \\
$p$ & image extent of the target object, px \\
$R$ & range from the sensor to the object, m \\
$\dot R$ & range rate, m\,s$^{-1}$ \\
$R_{\rm obj}$ & range from the sensor to the target object, m \\
$R_{\rm ref}$ & range from the sensor to the reference object, m \\
$S$ & physical size of the target object, m \\
$\hat{\mathbf{u}}$ & unit vector along the relative velocity \\
$\vobj$ & velocity of the object over the ground, m\,s$^{-1}$ \\
$\vown$ & velocity of the observing platform over the ground, m\,s$^{-1}$ \\
$\vpx$ & rate of the object's image across the frame, px\,frame$^{-1}$ \\
$W$ & frame width, px \\
$x$ & pixel offset of a feature from the center of the frame, px \\
$\alpha$ & field angle, i.e. angle between the camera axis and the direction to a
feature, rad \\
$\FOV$ & horizontal field of view of the image, rad \\
$\theta$ & aspect angle between the relative velocity vector and the viewing
direction, rad \\
$\Phi$ & angular size subtended by the target object, rad \\
$\omega$ & angular rate of the line of sight, rad\,s$^{-1}$ \\
\end{longtable*}}

\addtocounter{table}{-1}

\section{Introduction}\label{sec:intro}

Between late 2025 and mid-2026 the U.S.\ Department of War released five tranches of UAP records under the PURSUE initiative (\href{https://www.war.gov/UFO}{war.gov/UFO}). This included 112 sensor videos of targeting-pod and other turret mounted cameras (including infrared and electro-optical, meaning an electronic camera capturing images at optical wavelengths) from military aircraft \cite{pursue}; hyperlinks to the individual records are given in the Appendix. Motivated by the PURSUE release, this scientific analysis examines the extent to which an object's velocity can be constrained from these sensor videos.

As a known cautionary precedent, the U.S.\ Navy ``GOFAST'' video has been interpreted as showing an object at extreme speed near the sea surface; however, trigonometric reduction using its on-screen telemetry showed the apparent motion to be dominated by parallax from the sensor aircraft, with the object's implied speed consistent with wind-borne drift \cite{aaro}. In the GOFAST case, sufficient information was available to constrain the object's motion relative to the observer and assess it in relation to reported wind conditions.

This analysis shows that the PURSUE sensor videos provide even fewer constraints than were available in the GOFAST case. The paper first examines the quantities required to constrain the velocity of an object from a sensor video (Sec.~\ref{sec:framework}) and then summarizes the variables available in the PURSUE sensor videos (Sec.~\ref{sec:census}). Two candidates are then analyzed as a demonstration of the extent to which velocity can be constrained: DOW-UAP-PR113 (Sec.~\ref{sec:pr113}) includes visible depression-angle markings that provide a direct measurement of angular scale, and DOW-UAP-PR149 (Sec.~\ref{sec:pr149}) includes an object of known size that can be used as a scaling reference. The conclusion suggests that unredacted videos or corroborating data are needed to enable better constraints on velocity.

\section{The Velocity Equation}\label{sec:framework}

The kinematic quantities measurable from released pixels alone are the rate at which the object's image crosses the frame, $\vpx$ (pixels per frame), and the extent $p$ (pixels) that its image spans. The frame rate $f$ (frames per second) and frame size ($W \times H$ = width $\times$ height in pixels) are both read directly from the video container. These quantities can be converted to an angular rate of the line of sight ($\omega$ in $\text{rad\,s}^{-1}$) as
\begin{equation}\label{eq:omega}
\omega \;=\; \frac{\vpx\, f}{k},
\end{equation}
where $k$ is the sensor's angular scale (pixels per radian) at the object's position in the frame. Approaches for estimating the value of $k$ will be discussed in Sec.~\ref{sec:scale}. However, it is important to emphasize that $k$ is unconstrained for most of the PURSUE sensor videos (as discussed in Sec.~\ref{sec:census}), which makes $\omega$ itself unknown. Note that the All-domain Anomaly Resolution Office (AARO) analysis of GOFAST calculated $\omega$ by reading the pod's own azimuth and elevation off the on-screen display frame by frame. Such information is either redacted or unavailable in all of the PURSUE sensor videos.

\begin{figure}[hbt!]
\centering
\includegraphics[width=\textwidth]{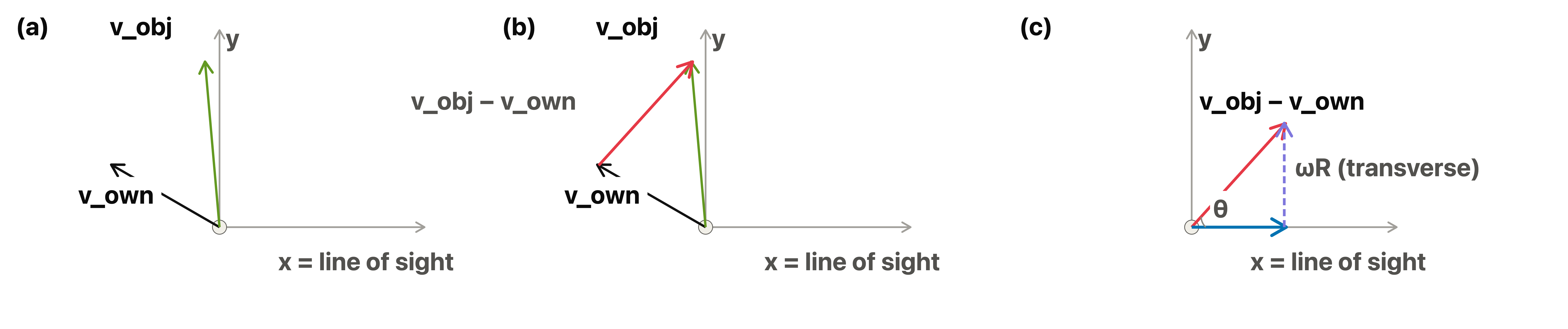}
\caption{Building the relative velocity decomposition used in Eq.~\eqref{eq:relation}, one step at a time. a) The aircraft's velocity, $\vown$, and the object's velocity, $\vobj$, drawn from the same starting point so they can be compared directly. b) Subtracting the two: $\vobj - \vown$ is the arrow that connects the tip of $\vown$ to the tip of $\vobj$. This is the relative velocity: how fast the object appears to move as seen from the aircraft. c) That same relative velocity arrow, moved to the origin and compared against the line of sight (the $x$-axis). Its component along the line of sight is the radial rate, $\dot R$; its component perpendicular to the line of sight is the transverse rate, $\omega R$. Only $\omega R$ shows up as a measurable pixel rate in the video. The value of $\dot R$, and therefore the object's true speed, cannot be recovered from a single sensor alone.}
\label{fig:diagram}
\end{figure}

The value of $\omega$ is generated by the combination of the object's motion and the aircraft's motion. For an object at range $R$ with velocity $\vobj$, observed from a platform with velocity $\vown$, the transverse component of the \emph{relative} velocity is what sweeps the line of sight (see the diagram in Fig.~\ref{fig:diagram}):
\begin{equation}\label{eq:relation}
\;\omega R \;=\; \big|\,\vobj-\vown\,\big|\,\sin\theta\;
\end{equation}
where $\theta$ is the aspect angle between the relative velocity vector and the viewing direction. It is important to note that $\theta$ is not a free parameter, given that the range itself changes with time. Accounting for this motion would require a range history sampled over time, whereas the PURSUE sensor videos do not include even a single reported range. The range rate, $\dot R = dR/dt$, is the quantity that a sensor could record, and $\theta$ is derived from the range rate as $\tan \theta=\omega R/|\dot R|$.

Solving for the object's velocity over the ground,
\begin{equation}\label{eq:solve}
\vobj \;=\; \vown \;+\; \frac{\omega R}{\sin\theta}\,\hat{\mathbf{u}},
\end{equation}
where $\hat{\mathbf{u}}$ is the unit direction of the relative velocity. In some of the sensor videos the image-plane projection of this relative velocity is measurable as $\vpx$, but expressing it in physical units still requires $k$ and $R$. However, obtaining the actual value of $\vobj$ would also require the velocity of the observing platform, $\vown$. Even in the GOFAST case, the trajectory of the aircraft was unknown, so the analysis could only report a relative velocity. Equations~\eqref{eq:omega}--\eqref{eq:solve} are the basis for the analysis that follows. The PURSUE sensor videos allow measurement of motion based on pixels ($\vpx$, $f$, $W$), but obtaining a physical velocity also requires four additional variables: $k$, $R$, $\dot R$ (or $\theta$ if $\dot R$ is unknown), and $\vown$.

\subsection{Object Size}\label{sec:size}

The angular scale of the sensor, $k$, also converts the object's image extent into a physical size. If the object's image spans $p$ pixels, then the angle it subtends is
\begin{equation}\label{eq:phi}
\Phi \;=\; \frac{p}{k},
\end{equation}
so that an object of angular size $\Phi$ at range $R$ has physical size $S = R\tan\Phi$. The images considered here span tens of pixels and therefore subtend $\Phi\lesssim1^\circ$. In this range, the tangent function is approximately linear, so the size of the object can be calculated using the small angle approximation, as
\begin{equation}\label{size_eq}
S \;=\; R\tan\Phi \approx  R\,\Phi \;= \; \frac{p\,R}{k}.
\end{equation}
Although the small angle approximation does not directly affect calculations of object size, uncertainties in the angular scale, $k$, can affect estimates of size.

\begin{figure}[hbt!]
\centering
\includegraphics[width=0.75\textwidth]{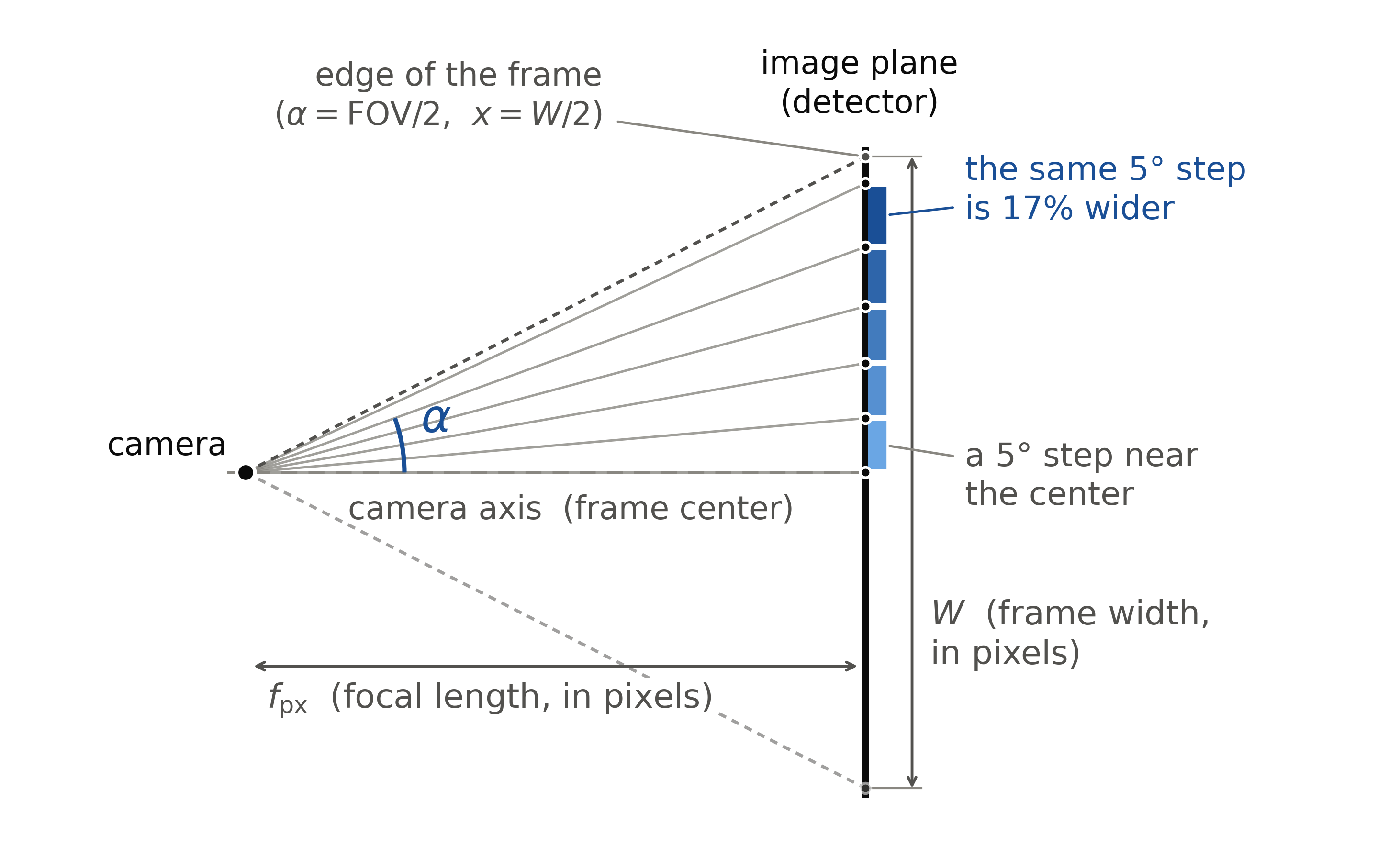}
\caption{The angular scale is not a constant value across an image frame. This example (drawn for $\FOV = 54^\circ$) shows that rays drawn from the camera in equal $5^\circ$ steps of field angle $\alpha$ are projected on the image plane at unequally spaced pixel offsets, because $x = f_{\rm px}\tan\alpha$. This makes steps near the frame edge wider than steps near the center. The detector spans the full frame width $W$, and the pixel offset $x$ is measured from the frame center, so the two edges of the frame lie at $x=\pm W/2$.}
\label{fig:scale}
\end{figure}

\subsection{Angular Scale}\label{sec:scale}

For a lens that renders straight lines in the world as straight lines in the image (known as a rectilinear lens), the pixel offset $x$ of a feature from the center of the frame is set by the angle between the camera axis and the direction to that feature, which is known as its field angle $\alpha$ (see the diagram in Fig.~\ref{fig:scale}):
\begin{equation}
  x = f_{\rm px}\tan\alpha,
\end{equation}
where the focal length $f_{\rm px}$ is expressed in pixels as
\begin{equation}
    f_{\rm px} = \frac{W/2}{\tan(\FOV/2)}.\label{eq:fpx}
\end{equation}
In Eq.~\eqref{eq:fpx}, $\FOV$ (rad) is the horizontal field of view in the image, with $\alpha=\FOV/2$ at the edge of the frame, and $W$ (pixels) is the frame width, with $x=W/2$ at the edge of the frame. Note that $\alpha$ indicates the position \textit{in the frame} that a feature appears; this is distinct from the aspect angle $\theta$ of Eq.~\eqref{eq:relation}, which describes the direction of an object's motion.

The angular scale of the sensor, $k=dx/d\alpha$, describes the number of pixels that a feature in the image shifts for a small change in field angle. This makes $k$ a function of $\alpha$ (position in the image frame) rather than a constant:
\begin{equation}\label{eq:scale}
k \;= \frac{dx}{d\alpha} = \; f_{\rm px}\sec^{2}\alpha \;=\; \frac{W/2}{\tan(\FOV/2)}\,\sec^{2}\alpha .
\end{equation}
This is the full expression for $k$, without making a small angle approximation.

We can consider a small angle approximation by noting that $k \to f_{\rm px}$ at the center of the frame ($\alpha=0$). Likewise, if the field of view is narrow, then $\tan(\FOV/2) \to \FOV/2$ and $k \to W/\FOV$. This gives the small angle approximation to Eq.~\eqref{eq:omega}:
\begin{equation}\label{eq:omegasmallangle}
\omega \;=\; \vpx\, f\frac{\FOV}{W}.
\end{equation}
However, this limit may not be applicable at the wide imaging fields that are typical of targeting pod cameras. For example, at $\FOV = 54^\circ$, the small angle approximation from Eq.~\eqref{eq:omegasmallangle} overstates the scale at the center of the frame by 8\% compared to Eq.~\eqref{eq:scale}. Furthermore, the full expression in Eq.~\eqref{eq:scale} shows that $k$ itself varies by 34\% between the center of the frame and its corner, which makes the small angle approximation of limited use in such cases.

In reality, constraints on $k$ for the PURSUE cases are based on the limited measurements or inferences that can be made from the redacted images. A clip that reports no scale, and includes no features available for scaling, is entirely unconstrained. A clip that reports at least one value for scale can at least be constrained near the part of the frame where the particular scale applies.

Two possible routes exist toward resolving this uncertainty in $k$. The first approach is to use a scale, report, or reconstructed value of $k$ directly or $\FOV$ (thus yielding $k$ through Eq.~\eqref{eq:scale}). This still carries uncertainties with the variation across the frame, and ultimately use of the small angle approximation (Eq.~\eqref{eq:omegasmallangle}) may be necessary to make any inferences at all. We illustrate this first approach in Sec.~\ref{sec:pr113}.

The second approach is to use an object of known physical size sharing the frame with the target, which makes $k$ cancel between the two measurements and gives a relative speed without the scale ever being determined. If the image frame contains an object with a known reference length $L$ that spans $l_{\rm px}$ pixels in the image, then Eq.~\eqref{size_eq} gives $k=l_{\rm px} R_{\rm ref}/L$, where $R_{\rm ref}$ is the range to the reference object. Substituting this into Eqs.~\eqref{eq:omega} and~\eqref{eq:relation} gives the transverse component of the contact's relative velocity as
\begin{equation}\label{eq:pr149}
\big|\,\vobj-\vown\,\big|\sin\theta \;=\;\vpx\, f \,\frac{L}{l_{\rm px}}\,\frac{R_{\rm obj}}{R_{\rm ref}} .
\end{equation}
Here the angular scale $k$ has canceled. Because both $\vpx$ and $l_{\rm px}$ are measured in the same image, this means no $\FOV$ is required. The relative velocity thus depends on the range ratio between the object and the reference, $R_{\rm obj}/R_{\rm ref}$. We illustrate this second approach in Sec.~\ref{sec:pr149}.

\section{The PURSUE Sensor Videos}\label{sec:census}

The 112 sensor videos were individually reviewed, each assessed for the availability of every variable in Eqs.~\eqref{eq:omega}--\eqref{eq:solve}. Each video was also categorized according to its kinematic properties (summarized in Table~\ref{tab:kin} and listed in full in the Appendix). A video is classified as a ``transit'' if it contains an object that crosses the field of view and moves against the background scene. A video is classified as ``tracked'' if it contains an object that stays inside the field of view, often oscillating about the center with properties that may indicate jitter as the sensor attempts to keep a lock on the object. A video is classified as ``both'' if it includes segments showing at least one transit and other segments with a tracked object. Each video is also classified according to whether it contains one or multiple objects.

\begin{table}[hbt!]
\caption{\label{tab:kin} Kinematic sort of the 112 sensor videos}
\centering
\begin{tabular}{lrrr}
\hline
Field-of-view behavior & Clips & One object & Multiple objects \\
\hline
Transit (crosses the field of view) & 43 & 34 & 9 \\
Tracked (held within the field of view) & 33 & 26 & 7 \\
Both (transits and is tracked at different times) & 36 & 19 & 17 \\
\hline
Total & 112 & 79 & 33 \\
\hline
\end{tabular}
\end{table}

No video contains complete information to constrain the relative velocity. The frame constants $f$ and $W$ are present in every file, but the angular scale $k$ and $\FOV$ are unreported. The rate of the object across the frame $\vpx$ is measurable only for transit events, as tracked objects that remain in the center of the frame show $\vpx\!\approx\!0$ by design. This near-zero rate is a property of the sensor rather than of the object: the pod slews to null the object's image motion, so $\vpx\!\approx\!0$ arises regardless of $\theta$. The transverse term $\omega R$ is therefore unrecoverable for tracked objects, and the relative motion $|\vobj-\vown|$ cannot be constrained for any of them. Observer velocity, $\vown$, and the aspect angle, $\theta$, (derived from the range rate $\dot R$) are absent from all 112 videos.

In summary, all videos fall short of providing sufficient information to place meaningful constraints on velocity. A single camera records angles alone, so neither $\dot R$ nor $\vown$ can be recovered from the imagery by any amount of analytical effort. This underscores a longstanding problem, identified by James McDonald, that single-sensor angular-only data cannot close kinematic questions, with multi-sensor observation being the only way to constrain the kinematic properties of anomalous objects \cite{mcdonald}. This multi-sensor and multi-site experimental approach is currently being developed by the Harvard-based Galileo Project \cite{loeb2023overview} and by the UAPx team \cite{matt2025}.

\section{Example Case: DOW-UAP-PR113}\label{sec:pr113}

\begin{figure}[hbt!]
\centering
\includegraphics[width=\textwidth]{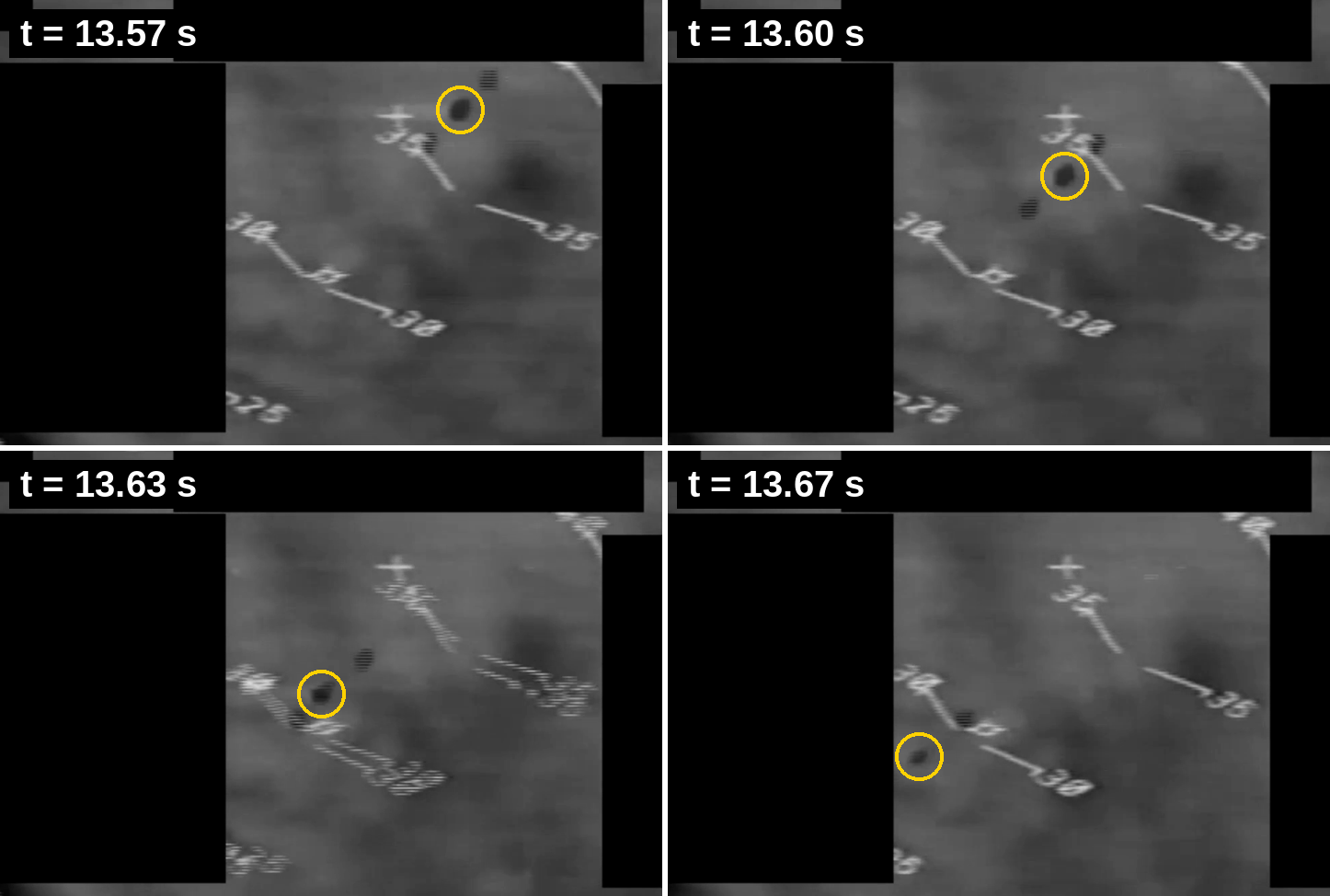}
\caption{The DOW-UAP-PR113 transit shown as four cropped and enlarged frames. The compact object (circled) crosses from upper center to lower left at 142\,px/frame in about $0.1$\,s. The bright diagonal ticks are the burned-in depression-angle graticule (labeled $5$--$35$) from which the field of view is recovered. Black rectangles are release redactions. The appearance of secondary images is an artifact due to video interlacing in the original footage and does not indicate multiple objects.}
\label{fig:cases}
\end{figure}

The infrared sensor video DOW-UAP-PR113 (Western United States, 1996, frame resolution $W\times  H=1920\times1080$, pixels at $f=30$\,frames per second) shows a compact dark body that crosses the frame at a measured $\vpx = 142$\,px/frame ($\approx$2.2 frame-widths per second), directly visible for only $\approx$0.1\,s across four progressive frames (Fig.~\ref{fig:cases}). The trajectory is linear to measurement precision and matches the release caption's description: ``enters near the upper right corner and exits near the lower left corner.''

This case is unique in the PURSUE corpus because DOW-UAP-PR113 carries a burned-in angular graticule labeled $5$--$35$ at 35.4\,px per unit (Fig.~\ref{fig:cases}, bright diagonal ticks). This appears to be a depression-angle scale in degrees, and it therefore measures the angular scale of Eq.~\eqref{eq:omega} directly, $k = 35.4$\,px\,deg$^{-1} = 2.03\times10^{3}$\,px\,rad$^{-1}$. With a linear extrapolation across the frame, this corresponds to $\FOV\approx54^\circ$ horizontal. Both of these quantities are derived from image analysis rather than released data, and both rest on the stated assumption that the graticule is a depression-angle scale in degrees. At this scale the measured $\vpx$ corresponds to $\omega \approx 2.10$\,rad\,s$^{-1}$ ($\approx$120$^\circ$\,s$^{-1}$).

\begin{figure}[hbt!]
\centering
\includegraphics[width=\textwidth]{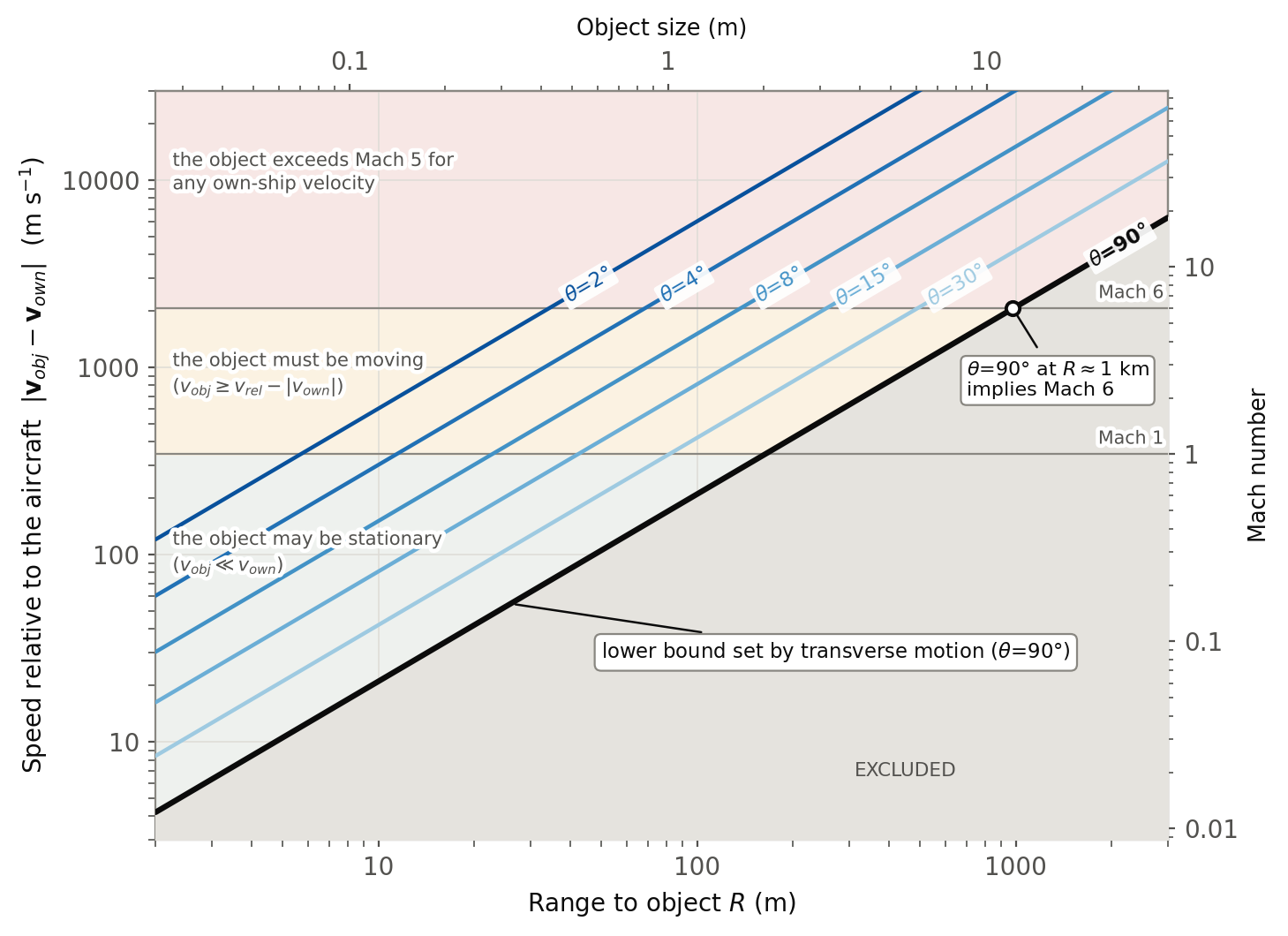}
\caption{Velocity constraints for DOW-UAP-PR113 based on available variables ($\vpx=142$\,px/frame, $f=30$\,fps, and $k=2.03\times10^{3}$\,px\,rad$^{-1}$ read from the graticule, which give $\omega=2.10$\,rad\,s$^{-1}$). The vertical axis shows the object's speed relative to the aircraft, $|\vobj-\vown| = \omega R/\sin\theta$, and the horizontal axis shows the unknown range $R$. Diagonal lines show various values of the unknown aspect angle $\theta$. Transverse motion ($\theta=90^\circ$, heavy black) is a hard lower bound, with the shaded region beneath excluded. The upper axis converts each range into the object size it implies from the measured $\sim$25\,px image extent using Eq.~\eqref{size_eq}.}
\label{fig:fan}
\end{figure}

DOW-UAP-PR113 is the only clip in which $\omega$ is constrained, but the object's physical velocity remains undetermined because $R$ and $\theta$ are unknown, and $\vown$ is not available. Figure~\ref{fig:fan} illustrates the limited insight available from these data, which shows the object's speed relative to the aircraft, $|\vobj-\vown| = \omega R/\sin\theta$, as a function of the unknown range $R$ and aspect angle $\theta$. Even this spans orders of magnitude, with subsonic velocities for a near-field object but supersonic velocities for objects beyond a few hundred meters for every aspect angle. The same ambiguity applies to the size of the object itself, with its measured $\sim$25\,px image extent consistent with anything from a bird flying past the sensor at close range to a $\sim$37\,m craft at 3\,km (upper axis of Fig.~\ref{fig:fan}).

\section{Example Case: DOW-UAP-PR149}\label{sec:pr149}

\begin{figure}[hbt!]
\centering
\includegraphics[width=\textwidth]{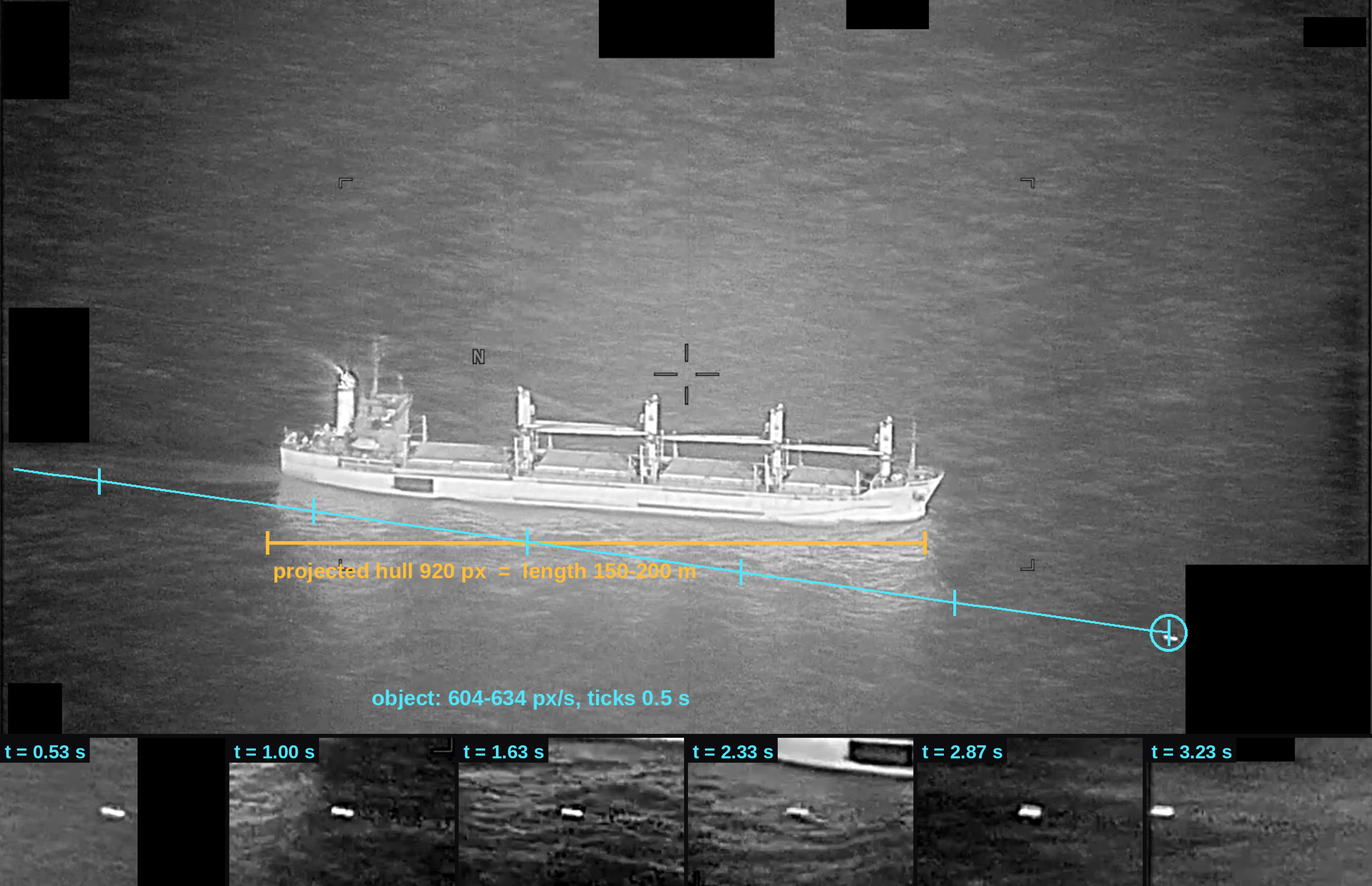}
\caption{The DOW-UAP-PR149 transit, shown with (top) the full frame at the start of the crossing, carrying the measured track with $0.5$\,s ticks and the ship's 920\,px projected hull marked as the scale bar on which the analysis rests, and (bottom) six crops centered on the object at $3\times$ magnification.}
\label{fig:pr149}
\end{figure}

The infrared sensor video DOW-UAP-PR149 (Middle East, 2023, frame resolution $W\times  H=1920\times1028$ at $f=30$\,frames per second)
opens with a compact object crossing the frame from the lower right toward the lower left at $\vpx=20.2$\,px/frame, tracked over 43 frames spanning $3.2$\,s on a path straight to $2.6$\,px rms (Fig.~\ref{fig:pr149}). The frames also show a carrier ship traveling beneath the object. The trajectory of the clipped segment matches the release caption's description: ``an area of contrast transits the frame from the bottom right edge to the lower left edge of the sensor field-of-view.''

Analysis of this clip circumvents the need to obtain the angular scale because the carrier ship can serve as a reference object of known physical size, which can then give relative velocity in terms of the range ratio between the object and the ship (following Eq.~\eqref{eq:pr149}). The scaling used is the ship's projected hull of $l_{\rm px}=920$\,px. A carrier ship of this class is approximately $L\approx150-200$\,m in length, corresponding to a ``Handymax'' class vessel \cite{manbulk}. The object moves through $0.657$--$0.689$ ship-lengths per second, crossing the entire hull in under $1.6$\,s. This gives an upper bound on the relative velocity for this case as
\begin{equation}\label{eq:pr149-upper}
\big|\,\vobj-\vown\,\big|\sin\theta \;\le\; (99-132\,\text{m s}^{-1})\,\frac{R_{\rm obj}}{R_{\rm ref}},
\end{equation}
where the velocity range corresponds to $\approx$\,0.3--0.4 Mach. The range ratio $R_{\rm obj}/R_{\rm ref}$ remains unknown, so Eq.~\eqref{eq:pr149-upper} can only provide an upper limit. The maximum relative velocity corresponds to the object and ship residing at the same distance from the sensor ($R_{\rm obj}/R_{\rm ref} = 1$). If this object is at an altitude higher than the ship, then the relative velocity will be less than this limit.

The range ratio remains open for DOW-UAP-PR149. This also only provides plausible bounds for the size limits of the object, which could include a $0.08$\,m body at $12$\,m\,s$^{-1}$ or a $0.15$\,m body at $22.5$\,m\,s$^{-1}$. In this case, supersonic relative velocities can be excluded, but the velocity and identity of the object cannot be further constrained.

\section{Discussion and Conclusions}\label{sec:discussion}

The PURSUE sensor videos can neither demonstrate nor exclude anomalous kinematics. Cases with a visible prosaic explanation can be set aside, but the remainder cannot be resolved unless additional data become available. Useful data that could help to resolve cases include: (i) the velocity of the platform ($\vown$) such as provided by flight logs; (ii) sensor metadata describing field-of-view and line-of-sight pointing angles; and (iii) range records or a correlated radar track. Not all of these data will exist for passive encounters, but releasing whatever does exist for a given case would help determine whether the objects in that video exhibit anomalous motion.

The GOFAST case was resolvable because the released sensor video included all on-screen sensor information without redactions. Other cases analyzed by AARO are resolvable because AARO has been granted access to classified information, which enables such analyses to draw on this sensor display data and other available sources. The PURSUE sensor videos have most or all of this information redacted, which implies that case resolution at the standards of GOFAST and other AARO analyses remains impossible without additional data.

This also illustrates an important distinction between disclosing a conclusion and releasing the evidence that supports it. An official announcement may be important and may prompt further study, but it does not by itself resolve a scientific question. Historically, scientific advances have not come from disclosure alone, but from observations and experiments that can be independently examined, reproduced or verified where possible, and evaluated through peer review. A sudden disclosure may begin this process but cannot replace it. The same standard should apply to UAP cases. An official conclusion becomes scientifically meaningful only when accompanied by sufficient data and methods for independent analysis; without them, it cannot be tested or used to place reliable constraints on the event.

In the absence of further data, attempts to derive kinematic limits from the PURSUE video corpus are unlikely to yield useful results. The sensor videos may retain value elsewhere, such as in the morphology of the reported objects, in the phenomenology of sensor artifacts, and in sharpening the data requests needed to resolve individual cases.

\section*{Appendix: Catalog of Sensor Videos}\label{sec:appendix}

Table~\ref{tab:cases} lists all 112 sensor videos in the PURSUE releases with the two
image-frame properties the corpus does determine for every clip, the sensor modality, and
the tranche in which each was released. Identifiers link to the DVIDS record.

\begingroup
\singlespacing
\footnotesize
\setlength{\LTcapwidth}{\textwidth}
\begin{longtable}{llccc}
\caption{The 112 released PURSUE sensor videos}\label{tab:cases}\\
\hline
Identifier & Sensor & Tranche & Behavior & Number \\
\hline
\endfirsthead
\multicolumn{5}{@{}l}{\footnotesize\textit{Table \ref{tab:cases} (continued)}}\\
\hline
Identifier & Sensor & Tranche & Behavior & Number \\
\hline
\endhead
\hline
\multicolumn{5}{@{}r}{\footnotesize\textit{(continued on next page)}}\\
\endfoot
\hline
\endlastfoot
\href{https://www.dvidshub.net/video/1017801}{FBI-UAP-PR007} & Infrared & 5 & Tracked & Multiple \\
\href{https://www.dvidshub.net/video/1006056}{DOW-UAP-PR19} & Infrared & 1 & Transit & One \\
\href{https://www.dvidshub.net/video/1006059}{DOW-UAP-PR21} & Infrared & 1 & Tracked & Multiple \\
\href{https://www.dvidshub.net/video/1006060}{DOW-UAP-PR22} & Infrared & 1 & Transit & One \\
\href{https://www.dvidshub.net/video/1006062}{DOW-UAP-PR23} & Infrared & 1 & Transit & One \\
\href{https://www.dvidshub.net/video/1014100}{DOW-UAP-PR024} & Infrared & 4 & Both & Multiple \\
\href{https://www.dvidshub.net/video/1006063}{DOW-UAP-PR26} & Infrared & 1 & Both & One \\
\href{https://www.dvidshub.net/video/1006067}{DOW-UAP-PR27} & Infrared & 1 & Both & One \\
\href{https://www.dvidshub.net/video/1006073}{DOW-UAP-PR28} & Infrared & 1 & Both & One \\
\href{https://www.dvidshub.net/video/1006074}{DOW-UAP-PR29} & Infrared & 1 & Tracked & One \\
\href{https://www.dvidshub.net/video/1014102}{DOW-UAP-PR030} & Infrared & 4 & Transit & Multiple \\
\href{https://www.dvidshub.net/video/1006076}{DOW-UAP-PR31} & Infrared & 1 & Transit & One \\
\href{https://www.dvidshub.net/video/1006078}{DOW-UAP-PR32} & Infrared & 1 & Tracked & One \\
\href{https://www.dvidshub.net/video/1006079}{DOW-UAP-PR33} & Infrared & 1 & Tracked & Multiple \\
\href{https://www.dvidshub.net/video/1006080}{DOW-UAP-PR34} & Infrared & 1 & Both & One \\
\href{https://www.dvidshub.net/video/1006082}{DOW-UAP-PR35} & Infrared & 1 & Transit & One \\
\href{https://www.dvidshub.net/video/1006083}{DOW-UAP-PR36} & Infrared & 1 & Transit & One \\
\href{https://www.dvidshub.net/video/1006087}{DOW-UAP-PR37} & Infrared & 1 & Transit & One \\
\href{https://www.dvidshub.net/video/1006088}{DOW-UAP-PR38} & Infrared & 1 & Transit & One \\
\href{https://www.dvidshub.net/video/1006089}{DOW-UAP-PR39} & Infrared & 1 & Transit & One \\
\href{https://www.dvidshub.net/video/1006093}{DOW-UAP-PR40} & Infrared & 1 & Transit & One \\
\href{https://www.dvidshub.net/video/1006094}{DOW-UAP-PR41} & Infrared & 1 & Tracked & One \\
\href{https://www.dvidshub.net/video/1006097}{DOW-UAP-PR42} & Infrared & 1 & Both & Multiple \\
\href{https://www.dvidshub.net/video/1006159}{DOW-UAP-PR43} & Infrared & 1 & Transit & One \\
\href{https://www.dvidshub.net/video/1006104}{DOW-UAP-PR44} & Infrared & 1 & Tracked & One \\
\href{https://www.dvidshub.net/video/1006105}{DOW-UAP-PR45} & Infrared & 1 & Both & One \\
\href{https://www.dvidshub.net/video/1006106}{DOW-UAP-PR46} & Infrared & 1 & Tracked & One \\
\href{https://www.dvidshub.net/video/1006107}{DOW-UAP-PR47} & Infrared & 1 & Tracked & Multiple \\
\href{https://www.dvidshub.net/video/1006110}{DOW-UAP-PR48} & Infrared & 1 & Tracked & One \\
\href{https://www.dvidshub.net/video/1006111}{DOW-UAP-PR49} & Electro-optical & 1 & Tracked & Multiple \\
\href{https://www.dvidshub.net/video/1007706}{DOW-UAP-PR050} & Infrared and electro-optical & 2 & Transit & Multiple \\
\href{https://www.dvidshub.net/video/1007707}{DOW-UAP-PR051} & Infrared and electro-optical & 2 & Both & One \\
\href{https://www.dvidshub.net/video/1007708}{DOW-UAP-PR052} & Infrared and electro-optical & 2 & Transit & Multiple \\
\href{https://www.dvidshub.net/video/1007709}{DOW-UAP-PR053} & Infrared & 2 & Transit & One \\
\href{https://www.dvidshub.net/video/1007711}{DOW-UAP-PR054} & Infrared & 2 & Tracked & One \\
\href{https://www.dvidshub.net/video/1007713}{DOW-UAP-PR055} & Infrared & 2 & Transit & One \\
\href{https://www.dvidshub.net/video/1007718}{DOW-UAP-PR056} & Infrared and electro-optical & 2 & Tracked & One \\
\href{https://www.dvidshub.net/video/1007720}{DOW-UAP-PR057a} & Infrared & 2 & Transit & One \\
\href{https://www.dvidshub.net/video/1007723}{DOW-UAP-PR058} & Infrared and electro-optical & 2 & Tracked & One \\
\href{https://www.dvidshub.net/video/1007727}{DOW-UAP-PR059} & Infrared & 2 & Both & One \\
\href{https://www.dvidshub.net/video/1007734}{DOW-UAP-PR060} & Infrared and electro-optical & 2 & Transit & One \\
\href{https://www.dvidshub.net/video/1007735}{DOW-UAP-PR061} & Infrared and electro-optical & 2 & Transit & One \\
\href{https://www.dvidshub.net/video/1007739}{DOW-UAP-PR062} & Infrared and electro-optical & 2 & Transit & One \\
\href{https://www.dvidshub.net/video/1007740}{DOW-UAP-PR063} & Infrared and electro-optical & 2 & Transit & One \\
\href{https://www.dvidshub.net/video/1007741}{DOW-UAP-PR064} & Infrared & 2 & Transit & One \\
\href{https://www.dvidshub.net/video/1007777}{DOW-UAP-PR065} & Electro-optical & 2 & Transit & One \\
\href{https://www.dvidshub.net/video/1007778}{DOW-UAP-PR066} & Infrared & 2 & Transit & One \\
\href{https://www.dvidshub.net/video/1007779}{DOW-UAP-PR067} & Infrared & 2 & Transit & Multiple \\
\href{https://www.dvidshub.net/video/1007780}{DOW-UAP-PR068} & Infrared & 2 & Tracked & Multiple \\
\href{https://www.dvidshub.net/video/1007781}{DOW-UAP-PR069} & Infrared and electro-optical & 2 & Both & One \\
\href{https://www.dvidshub.net/video/1007783}{DOW-UAP-PR070} & Infrared & 2 & Tracked & One \\
\href{https://www.dvidshub.net/video/1007784}{DOW-UAP-PR071} & Infrared & 2 & Tracked & One \\
\href{https://www.dvidshub.net/video/1007788}{DOW-UAP-PR072} & Electro-optical & 2 & Tracked & One \\
\href{https://www.dvidshub.net/video/1007790}{DOW-UAP-PR073} & Infrared & 2 & Both & Multiple \\
\href{https://www.dvidshub.net/video/1007791}{DOW-UAP-PR074} & Infrared and electro-optical & 2 & Transit & Multiple \\
\href{https://www.dvidshub.net/video/1007795}{DOW-UAP-PR075} & Infrared & 2 & Tracked & One \\
\href{https://www.dvidshub.net/video/1007804}{DOW-UAP-PR076} & Infrared & 2 & Both & One \\
\href{https://www.dvidshub.net/video/1007809}{DOW-UAP-PR077} & ATFLIR, channel unstated & 2 & Both & Multiple \\
\href{https://www.dvidshub.net/video/1007812}{DOW-UAP-PR078} & ATFLIR, channel unstated & 2 & Both & Multiple \\
\href{https://www.dvidshub.net/video/1007816}{DOW-UAP-PR079} & Infrared and electro-optical & 2 & Tracked & Multiple \\
\href{https://www.dvidshub.net/video/1007803}{DOW-UAP-PR080} & Infrared & 2 & Transit & Multiple \\
\href{https://www.dvidshub.net/video/1007805}{DOW-UAP-PR081} & Infrared & 2 & Transit & One \\
\href{https://www.dvidshub.net/video/1007807}{DOW-UAP-PR082} & Infrared & 2 & Transit & One \\
\href{https://www.dvidshub.net/video/1007808}{DOW-UAP-PR083} & Infrared and electro-optical & 2 & Both & Multiple \\
\href{https://www.dvidshub.net/video/1007810}{DOW-UAP-PR084} & Infrared & 2 & Transit & One \\
\href{https://www.dvidshub.net/video/1007796}{DOW-UAP-PR085} & ATFLIR, channel unstated & 2 & Transit & One \\
\href{https://www.dvidshub.net/video/1007797}{DOW-UAP-PR086} & Electro-optical & 2 & Transit & One \\
\href{https://www.dvidshub.net/video/1007799}{DOW-UAP-PR087} & Infrared & 2 & Transit & Multiple \\
\href{https://www.dvidshub.net/video/1007800}{DOW-UAP-PR088} & ATFLIR, channel unstated & 2 & Both & Multiple \\
\href{https://www.dvidshub.net/video/1007712}{DOW-UAP-PR089} & ATFLIR, channel unstated & 2 & Both & Multiple \\
\href{https://www.dvidshub.net/video/1007719}{DOW-UAP-PR090} & Infrared & 2 & Transit & One \\
\href{https://www.dvidshub.net/video/1007716}{DOW-UAP-PR091} & Infrared & 2 & Both & Multiple \\
\href{https://www.dvidshub.net/video/1007715}{DOW-UAP-PR092} & Infrared & 2 & Both & One \\
\href{https://www.dvidshub.net/video/1007721}{DOW-UAP-PR093} & ATFLIR, channel unstated & 2 & Transit & Multiple \\
\href{https://www.dvidshub.net/video/1007722}{DOW-UAP-PR094} & Infrared and electro-optical & 2 & Both & One \\
\href{https://www.dvidshub.net/video/1007725}{DOW-UAP-PR095} & ATFLIR, channel unstated & 2 & Both & Multiple \\
\href{https://www.dvidshub.net/video/1007726}{DOW-UAP-PR096} & Infrared & 2 & Transit & Multiple \\
\href{https://www.dvidshub.net/video/1007728}{DOW-UAP-PR097} & Infrared & 2 & Both & Multiple \\
\href{https://www.dvidshub.net/video/1007737}{DOW-UAP-PR098} & Infrared & 2 & Both & Multiple \\
\href{https://www.dvidshub.net/video/1007738}{DOW-UAP-PR099} & Infrared & 2 & Both & Multiple \\
\href{https://www.dvidshub.net/video/1014096}{DOW-UAP-PR100} & Infrared & 4 & Transit & One \\
\href{https://www.dvidshub.net/video/1014097}{DOW-UAP-PR101} & Infrared & 4 & Both & Multiple \\
\href{https://www.dvidshub.net/video/1014098}{DOW-UAP-PR102} & Infrared & 4 & Both & One \\
\href{https://www.dvidshub.net/video/1014099}{DOW-UAP-PR103} & Infrared & 4 & Both & One \\
\href{https://www.dvidshub.net/video/1014101}{DOW-UAP-PR104} & Infrared & 4 & Tracked & One \\
\href{https://www.dvidshub.net/video/1014103}{DOW-UAP-PR105} & Infrared & 4 & Transit & One \\
\href{https://www.dvidshub.net/video/1014104}{DOW-UAP-PR106} & Electro-optical & 4 & Transit & One \\
\href{https://www.dvidshub.net/video/1014105}{DOW-UAP-PR107} & Infrared and electro-optical & 4 & Both & One \\
\href{https://www.dvidshub.net/video/1014106}{DOW-UAP-PR108} & Infrared & 4 & Both & One \\
\href{https://www.dvidshub.net/video/1014108}{DOW-UAP-PR109} & Infrared & 4 & Transit & One \\
\href{https://www.dvidshub.net/video/1014112}{DOW-UAP-PR110} & Infrared & 4 & Tracked & One \\
\href{https://www.dvidshub.net/video/1014114}{DOW-UAP-PR111} & ATFLIR, channel unstated & 4 & Both & Multiple \\
\href{https://www.dvidshub.net/video/1014128}{DOW-UAP-PR112} & Infrared & 4 & Both & One \\
\href{https://www.dvidshub.net/video/1014119}{DOW-UAP-PR113} & Infrared & 4 & Transit & One \\
\href{https://www.dvidshub.net/video/1014121}{DOW-UAP-PR114} & Infrared & 4 & Both & One \\
\href{https://www.dvidshub.net/video/1014123}{DOW-UAP-PR115} & Infrared & 4 & Both & One \\
\href{https://www.dvidshub.net/video/1014124}{DOW-UAP-PR116} & Infrared & 4 & Tracked & One \\
\href{https://www.dvidshub.net/video/1017793}{DOW-UAP-PR117} & Infrared & 5 & Tracked & One \\
\href{https://www.dvidshub.net/video/1017795}{DOW-UAP-PR118} & Infrared & 5 & Tracked & One \\
\href{https://www.dvidshub.net/video/1017798}{DOW-UAP-PR119} & Infrared & 5 & Tracked & One \\
\href{https://www.dvidshub.net/video/1017800}{DOW-UAP-PR120} & Infrared & 5 & Tracked & One \\
\href{https://www.dvidshub.net/video/1017802}{DOW-UAP-PR121} & Infrared & 5 & Both & Multiple \\
\href{https://www.dvidshub.net/video/1017803}{DOW-UAP-PR122} & Infrared & 5 & Both & Multiple \\
\href{https://www.dvidshub.net/video/1017805}{DOW-UAP-PR123} & Infrared & 5 & Tracked & One \\
\href{https://www.dvidshub.net/video/1017806}{DOW-UAP-PR124} & Infrared & 5 & Tracked & One \\
\href{https://www.dvidshub.net/video/1017788}{DOW-UAP-PR125} & Infrared & 5 & Tracked & One \\
\href{https://www.dvidshub.net/video/1017790}{DOW-UAP-PR126} & Electro-optical & 5 & Tracked & One \\
\href{https://www.dvidshub.net/video/1017791}{DOW-UAP-PR127} & Electro-optical & 5 & Tracked & One \\
\href{https://www.dvidshub.net/video/1017792}{DOW-UAP-PR134} & Infrared and electro-optical & 5 & Transit & One \\
\href{https://www.dvidshub.net/video/1017796}{DOW-UAP-PR136} & Infrared & 5 & Tracked & One \\
\href{https://www.dvidshub.net/video/1017797}{DOW-UAP-PR142} & Infrared & 5 & Transit & One \\
\href{https://www.dvidshub.net/video/1017799}{DOW-UAP-PR149} & Infrared & 5 & Both & One
\end{longtable}
\vspace{-\baselineskip}
\endgroup

\section*{Funding Sources}

This research did not receive any specific grant from funding agencies in the public, commercial, or not-for-profit sectors.

\section*{Acknowledgments}
Analysis tooling was developed with the assistance of Claude Opus 5 \cite{anthropic2026opus5}. Any opinions, findings, and conclusions or recommendations expressed in this material are those of the authors and do not necessarily reflect the views of any agency or employer.

\bibliography{refs}

\end{document}